# POSITIVE INFLUENCE OF Ta ADDITION ON SUPERCONDUCTIVE PROPERTIES OF HIGH PRESSURE SYNTHESIZED $MgB_2$

Tatiana PRIKHNA[(1)], Wolfgang GAWALEK[(2)], Nikolay NOVIKOV[(1)], Yaroslav SAVCHUK[(1)], Viktor MOSHCHIL[(1)], Nina SERGIENKO[(1)], Michael WENDT[(2)], Sergey DUB[(1)], Vladimir MELNIKOV[(3)], Alexey SURZHENKO[(2)], Doris LITZKENDORF[(2)], Petr NAGORNY[(1)], Christa SCHMIDT[(2)].

[(1)]Institute for Superhard Materials of the Nat'l Ac. Sci of Ukraine, 2, Avtozavodskaya St., 04074, Kiev, Ukraine
[(2)]Institut für Physikalische Hochtechnologie e.V., Winzerlaer Str. 10, D-07743 Jena, Germany
[(3)]Institute of Geochemistry, Mineralogy and Ore-Formation of the Nat'l Ac. Sci of Ukraine 34, Palladin Pr., 02142 Kiev, Ukraine

Bulk $MgB_2$ – based materials with the following critical current densities ($j_c$) in 1 T field: 570 $kA/cm^2$ at 10 K, 350 $kA/cm^2$ at 20 K and 40 $kA/cm^2$ at 30 K and in 10 T field: 650 $A/cm^2$ at 10 K have been high-pressure (HP) synthesized from Mg and B with 2 - 10 wt.% addition of Ta. In synthesis and sintering process Ta plays the role of an absorbent of impurity gases (hydrogen, nitrogen, etc.) and forms $Ta_2H$, $TaH$, $TaN_{0.1}$, etc., thus promotes the reduction of $MgH_2$ in Mg-B-O-matrix phase, as well as, the impurity nitrogen and oxygen in Mg-B grains (black $MgB_2$ single crystals distributed over the matrix). Vickers microhardness of HP-synthesized material is $H_v=12.54\pm0.86$ GPa (at 0.496-N load) The hardness ( at 60-mN load) of $MgB_2$ single crystals located in a sample matrix is $35.6\pm0.9$ GPa that is higher than the hardness of sapphire ($31.1\pm2.0$ GPa).

1. INTRODUCTION

Superconductive (SC) properties of $MgB_2$ have been revealed in January 2001. Since that time the impetuous progress in studying methods of preparation of $MgB_2$-based materials as well as the influence of different additions on their SC characteristics has been observed. The positive influences of a small amount of hydrogen if it is present in the structure of $MgB_2$ powder: the $MgB_2H_{0.03}$ powder has a little bit higher temperature of SC transition $T_c$ (by 0.5 K); of Zn doping, which increases $T_c$ by less than one degree, and of the substitution of oxygen for barium on critical current density $j_c$ have been reported by V.V. Flambaum et al.[1] , S.M. Kazakov et al.[2] and R.F. Klie et al.[3], respectively. R.F. Klie et al.[3] found ~20-100 nm precipitates that were formed by the ordered substitution of oxygen atoms onto boron lattice sites and that the basic bulk $MgB_2$ crystal structure and orientation were preserved. The periodicity of the oxygen ordering was dictated by the oxygen concentration in the precipitates and primarily occurred in the (010) plane. The presence of these precipitates correlated well with an improved $j_c$ and SC transition behavior, implying that they act as pinning centers.

Superconductive compounds in the Ta-B system are: $Ta_2B$ ($T_c$=3.12 K) and $TaB_2$ (some of the researchers consider that this compound has $T_c$=9.5 K but others report that it is not superconductive) [4].

It have been mentioned[4], that the existence of different critical temperatures for the starting $MgB_2$ at zero doping levels may give different Tc(x) behaviors (where x is the deviation of Mg content from the stoichiometry), and that Mg nonstoichiometry leads to different critical temperature dependencies on the applied pressure, therefore, one could expect the different Tc(x) behaviours as a function of small Mg nonstoichiometry[4].

We have established[5] that the $j_c$ of the synthesized material is positively affected by Ta present as a foil that covered samples or added to the starting Mg-B powdered mixture (2 wt.% of Ta powder). It should be noted that the positive influence of Ta on SC properties of high-pressure synthesized $MgB_2$ was observed for the samples manufactured using high $H_3BO_3$ impurity boron. $H_3BO_3$ have been formed as a result of long storing (for 25 years) of the amorphous boron (of about 95% purity). Here we present our results obtained for the HP-synthesized samples with 2 and 10 % of Ta addition using fresh prepared 95-97% purity amorphous B. We obtained further confirmations of our previous conclusion that Ta positively affect the $T_c$ and irreversible field $H_{irr}$ of the $MgB_2$-based bulk material. Our results on properties of high-pressure sintered $MgB_2$ (from $MgB_2$ powder) will be also discussed.

2. EXPERIMENTAL

In the experiments on synthesis, metallic Mg scobs and amorphous B (95-97 % purity) have been taken in the stoichiometric ratio of $MgB_2$. To study the influence of Ta, the Ta metallic powder has been added to the stoichiometric mixture of Mg and B in the amount of 2 and 10 wt. %. Then we mixed and milled the components in a high-speed activator with steel balls for 1-3 min. The obtained powder was compacted into tablets. For the experiments on sintering the commercial $MgB_2$ powder of Alfa Aesar company have been used.

High-pressure has been created inside a high-pressure apparatuses (HPA) of the recessed-anvil and cube (six punches) types, described elsewhere.[6] The working volume of the biggest cube-type HPA is of about 100 $cm^3$ (sample can be up to 60 mm in diameter). The sample was in contact with a compacted powder of hexagonal BN or enveloped in a Ta-foil and then placed inside the compacted monoclinic $ZrO_2$ powder Samples were synthesized and sintered under pressure 2 GPa for 1 h at 800, 900 and

950 °C: (1) without Ta addition in contact with BN and (2) with 2 and 10 wt.% of Ta addition in contact with Ta (i.e. enveloped in Ta-foil and placed into zirconia).

The structure of materials was studied using SEM and energy dispersive X-ray analysis, polarizing microscopy, and X-ray diffraction analysis. The $j_c$ was estimated from magnetization hysteresis loops obtained on an Oxford Instruments 3001 vibrating sample magnetometer (VSM) using Bean's model[7]. Hardness was measured on a Matsuzawa Mod. MXT-70 microhardness tester by a Vickers indenter. Nanohardness and Young modulus were examined using Nano Indenter-II, MTS Systems Corporation, Oak Ridge, TN, USA. The fracture toughness was estimated from the length of the radial cracks emanating from the corners of an indent.

## 3. RESULTS AND DISCUSSION

Fig.1 demonstrates the results of X-ray and VSM (the dependence of $j_c$ on magnetic field at different temperatures) study of HP-synthesized samples under different conditions. The highest $j_c$ in the 1-2 T field has been exhibited by the sample synthesized at 2 GPa, 900 °C, 1h with 2 wt. % of Ta addition (Fig.1d), while the highest $j_c$ in the 2-10 T field and irreversible field ($H_{irr}$), the sample, synthesized at 2 GPa, 800 °C, 1h with 10 wt. % of Ta addition (Fig. 1e). Fig.1g shows the X-ray pattern and $j_c$ vs. $\mu_o H_{irr}$ relations for the optimal (from the point of view of a compromise between the highest $j_c$ and $H_{irr}$) sample synthesized at 2 GPa, 800 °C, 1h with 2 wt. % of Ta addition. Analyzing the X-ray and VSM data we have come to the conclusion that samples with higher superconductive (SC) characteristics contain more unreacted Mg, and smaller amount or totally absent of $Ta_2H$ (both with orthorhombic and tetragonal structure). In the samples with Ta addition, the $Ta_2H$ phase has been observed and its amount increases with the increase Ta content of the sample. The residual unreacted Ta was found in a small amount in the sample with 10 % of Ta addition only (Fig. 1e).

In Fig.2 you can see the SEM (composition images) of the HP-synthesized and HP-sintered samples. They have a multiphase nanostructure. The black grains or single crystal inclusions (from the micron or even less to dozen microns in size) of Mg-B phase ($MgB_2$) are distributed over the matrix. It should be noted that the samples with higher $j_c$ and $H_{irr}$ have the higher density of Mg-B ($MgB_2$) inclusions in their matrix, i.e. the higher amount of black grains (see, for example, Figs. 2 c, d). The distribution of black $MgB_2$ grains over the matrix in a high-pressure synthesized material is more homogeneous

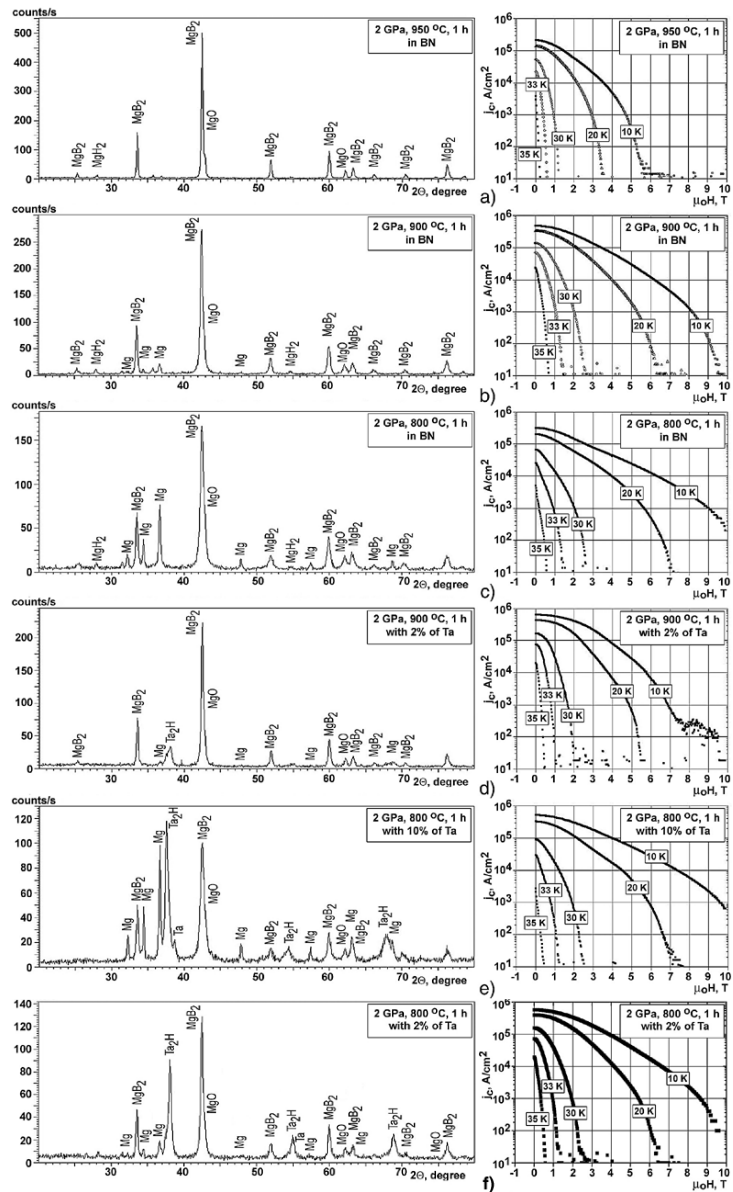

FIGURE 1

The X-ray patterns (left) and the dependence of critical current density $j_c$ on magnetic field $\mu_oH_{irr}$ under different temperatures (right) for the high-pressure synthesized samples. The synthesis conditions (pressure, in GPa, temperature, in °C, time, in h, amount of Ta addition, in wt.%, material in contact) are given in the upper right corner of the graphs.

than that in the sintered one (Fig.2 a, b) and it has better SC properties. The density of black grains in the sintered material is higher in the places of former boundaries between the initial particles, thus, black grains seem to "repeat" the structure of the initial powder. Fig.3 presents the energy dispersive spectra obtained by SEM that characterize the amount of elements in phases. The gray (matrix phase) of the samples consists mainly of Mg, B, O. Matrix phase of the sample with better SC properties contain higher amount of boron and a little bit less amount of oxygen (Fig. 3b). Black Mg-B ($MgB_2$) inclusions in

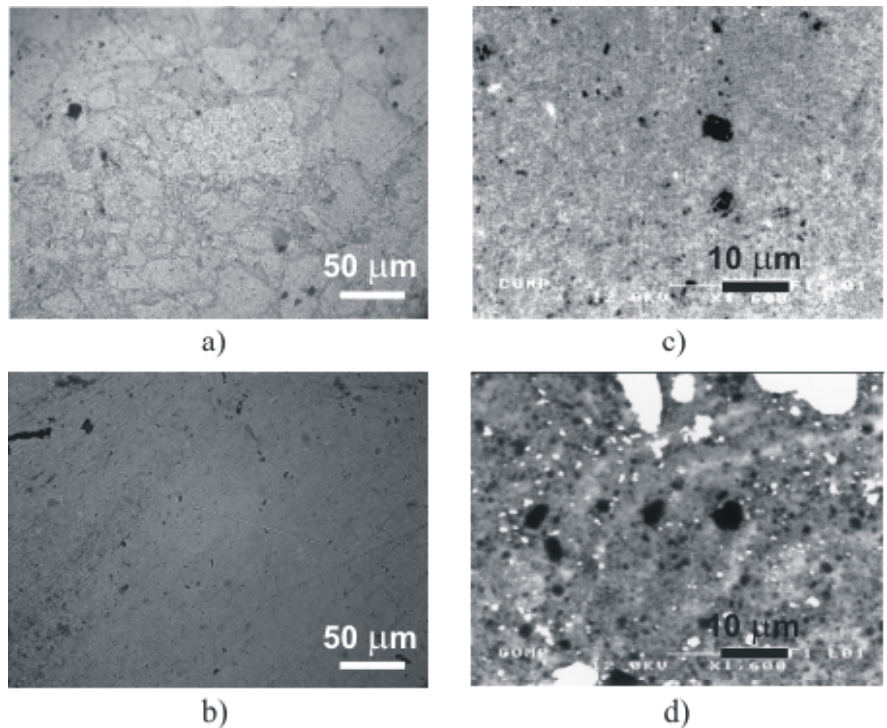

FIGURE 2

Structures (polarizing microscopy) of $MgB_2$ obtained at 2 GPa, 900 °C, 1 h:
(a) by sintering in BN of $MgB_2$ powder ( the $j_c$ in 1T at 20 K was 20 kA/cm$^2$) and
(b) by synthesis from Mg and B with 2 wt.% of Ta ( the $j_c$ in 1T at 20 K was 350 kA/cm$^2$)

SEM pictures (composition images) that show the different concentrations of black $MgB_2$ grains in the HP-synthesized samples at 2 GPa, 1h:
(c) at 950 °C in contact with BN ( the $j_c$ in 1T at 20 K was 73 kA/cm$^2$) and
(d) at 800 °C with 10 wt.% of Ta addition ( the $j_c$ in 1T at 20 K was 240 kA/cm$^2$).

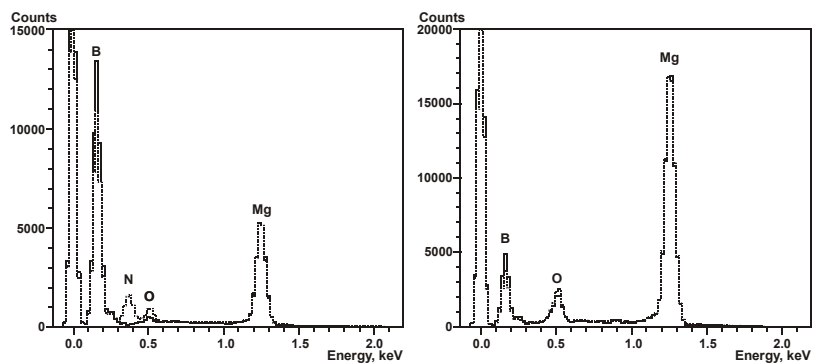

FIGURE 3

The energy dispersive spectra obtained by microprobe analysis (on SEM) that characterize the element contents of the samples synthesized under 2 GPa, 1 h:
(a) in the black grains (most likely, $MgB_2$ single crystalline inclusions into the matrix phase) of the same samples:
solid line – sample with 10 wt.% of Ta addition synthesized at 800 °C (see also Fig.1e)
dashed line – sample synthesized at 950 °C in contact with BN (see also Fig.1a)
(b) in the gray matrix phase (see also Fig.2 c, d) :
solid line – sample with 10 wt.% of Ta addition synthesized at 800 °C (see also Fig.1e)
dashed line – sample synthesized at 950 °C in contact with BN (see also Fig.1a)

Table 1. Microhardness, nanohardness, hardness, fracture toughness, and Young modulus of sintered and synthesized $MgB_2$ and a sapphire single crystal

| Characteristics | Matrix phase of the samples | Single crystal $MgB_2$ inclusions | Sapphire, $Al_2O_3$ |
|---|---|---|---|
| High –pressure sintered $MgB_2$ | | | |
| Indentation load: 60 mN (Berkovich indenter) | | | |
| Nanohardness, $H_B$, GPa | 17.4±1.1 | 35.6±0.9 | 31.1±2.0 |
| Young modulus, E, GPa | 213±18 | 385±14 | 416±22 |
| Indentation load: 4.96 N (Vickers indenter) | | | |
| Microhardness, $H_v$, GPa | 17.1±1.11 | - | - |
| Indentation load: 147.2 N (Vickers indenter) | | | |
| Fracture toughness, $K_{1c}$, MN·m$^{-3/2}$ | 7.6±2.0 | - | - |
| Hardness, $H_v$, GPa | 10.12±0.2 | - | - |
| High –pressure synthsised $MgB_2$ | | | |
| Indentation load: 0.496 N (Vickers indenter) | | | |
| Microhardness, $H_v$, GPa | 12.54±0.86 | - | - |

the HP-synthesized samples with Ta addition contain no impurity nitrogen and less impurity oxygen than those in the samples without Ta addition (Fig. 3a). Besides, these black Mg-B inclusions in the samples with better SC properties contain a higher amount of B than those in the samples with worse SC characteristics while the amount of Mg is about the same (Fig. 3a). In the process of $MgB_2$ sintering Ta turned into $TaN_{0.1}$.

The positive influence of Ta implies that in synthesis process it absorbed gases such as, hydrogen and nitrogen. We have never found any Ta-Mg, Ta-B or Ta-Mg-B compounds in the synthesized material. The presence of Ta allows us also to extend the temperature region of synthesis to obtain a material with high $j_c$ and $H_{irr}$. Positive influence of Ta is much more pronounced in the synthesis process than in the sintering one. Obtained by us HP-synthesized $MgB_2$- based material has highest data on $j_c$ and $H_{irr}$ reported up to now in literature for bulk $MgB_2$ (by H. P. Kim et al.[8] for sintered $MgB_2$).

Table 1 shows mechanical characteristics of sintered and synthesized samples under different indentation loads. In some cases we have to apply a very high indentation load, because, for example, in the case of fracture toughness there are no cracks from the corners of the indent mark under low loads, which rendered the study of this characteristic impossible. For the first time the microhardness of $MgB_2$ single crystals has been estimated and it has turned out to be higher than that for a sapphire single crystal.

4. CONCLUSIONS

Ta additions during synthesis of $MgB_2$ positively affect the material $j_c$ and $H_{irr}$ because of gases absorption: of hydrogen, nitrogen, etc. The structure of materials with higher $j_c$ and $H_{irr}$, contains a higher amount of black $MgB_2$ grains in the matrix phase; higher B, and lower impurity N and O concentrations in black $MgB_2$; $MgB_2$ grains are more uniformly distributed over the matrix; some amount of unreacted Mg, low amount (or absence) of $MgH_2$. The hardness of single-crystal $MgB_2$ is higher than that of a sapphire single crystal. HP-synthesized $MgB_2$-based material is promising for practical applications such as electromotors, flywheels, frictionless bearings, etc., its SC characteristics ($j_c$, $H_{irr}$) are higher than those ever reported.